\documentclass[12pt]{iopart}

\usepackage{graphicx}
\usepackage{bm}
\usepackage{times}
\usepackage{txfonts}
\usepackage{epstopdf}
\usepackage{latexsym}
\usepackage{epsfig}

\newcommand{\nbar}{\overline{n}}

\begin{document}

\title[Entanglement distribution in optomechanical devices]{Driven optomechanical systems for mechanical entanglement distribution}

\author{Mauro Paternostro, Laura Mazzola, and Jie Li}
\address{Centre for Theoretical Atomic, Molecular and Optical Physics, School of Mathematics and Physics, Queen's University Belfast, BT7 1NN Belfast, United Kingdom}
\ead{m.paternostro@qub.ac.uk}
\begin{abstract}
We consider the distribution of entanglement from a multi-mode optical driving source to a network of remote and independent optomechanical systems. By focusing on the tripartite case, we analyse the effects that the features of the optical input states have on the degree and sharing-structure of the distributed, fully mechanical, entanglement. This study, which is conducted looking at the mechanical steady-state, highlights the structure of the entanglement distributed among the nodes and determines the relative efficiency between bipartite and tripartite entanglement transfer. We discuss a few open points, some of which directed towards the bypassing of such limitations. 

\end{abstract}

\submitto{\JPB}
\maketitle

The interaction between mechanical devices and light has been the focus of a very intensive research activity, in the last six years, directed towards the demonstration of quantum control achieved and operated over quasi-macroscopic systems undergoing quite non-trivial dynamics~\cite{optoreview}. This has culminated in the very recent experimental demonstrations of ground-state cooling of massive mechanical modes achieved through radiation pressure-induced dynamical back-action~\cite{cooling1}, the realization of important technological steps directed towards the construction of quantum-limited position sensors and transducers~\cite{tech}, as well as the engineering of non-linearities strong enough to mimic, in a fully mechanical context, processes and dynamics that were so far typical of quantum optical settings~\cite{nonlinear}. The perspective of exploiting massive mechanical systems in order to prepare quantum states and enforce non-classical dynamics has spurred a few theoretical proposals for the achievement of entanglement involving both optical and mechanical systems (from here on referred to as {\it optomechanical entanglement} for brevity)~\cite{entanglement}. Such a scenario has been enlarged to consider schemes for the generation of fully mechanical entangled states~\cite{allmechanical,mazzolapatern} and quantum correlations in hybrid systems involving light, mechanical modes and collective atomic ones~\cite{ibrido}. Such achievements are endowed with intrinsic fundamental interest as they open up the way to the experimental study of mechanical versions of the long-craved {\it Schr\"odinger cat} and Einstein-Podolski-Rosen state, as well as the investigation of exotic decoherence mechanisms~\cite{deco} including the quantum gravitational collapse model~\cite{ellis}, the continuous spontaneous localization process~\cite{continuous} and gravitationally induced decoherence~\cite{gravity}. Moreover, the considerable flexibility and high potential for hybridization of optomechanical systems leave sufficient space for a significant re-design of the way a device for quantum communication is thought of. Taking inspiration from the (by now well grounded) paradigm of quantum networks for communication made out of local processing nodes, mutually interconnected by quantum channels~\cite{kimble,paternostrovari}, one can think about a web of remote and non-interacting optomechanical systems that share multi-photon light fields ensuring long-haul connectivity. In a sense, the embryo of such an architecture was already in the work by Zhang {\it et al.}~\cite{braunstein}, where the coupling between a two-mode squeezed vacuum (TMSV) state and two independent optomechanical cavities was explored for mechanical Hamiltonian engineering. This scheme has been pushed further in~\cite{mazzolapatern}, where an extensive study of the efficiency of entanglement distribution from purely optical TMSV resources to all-mechanical receivers, including the proposal of a mechanism for the measurement of such entanglement, was presented. It should be noticed that alternative configurations for the distribution of entanglement in a mechanical network itself have been proposed, based on optical postselection~\cite{borkje} or the use of atomic information carriers propagating in close proximity of monolithic microresonators~\cite{aoki}. 

In this paper, we go significantly beyond the {\it proof-of-principle} study reported in~\cite{mazzolapatern} and analyze the performance of mechanical entanglement distribution by means of more general multimode optical resources to determine the extents of the validity of such paradigm. We start by digging deep into the two-mode resource case, which entails the minimal configuration for entanglement distribution and, as such, the basic building block for any optomechanical network. By working with generally mixed two-mode Gaussian states, we show that the mechanism is very much relying on the symmetry of the optical resource's state: small asymmetries in the statistics of the two modes composing the optical entanglement carrier correspond to a spoiled mechanical entanglement distribution efficiency. The distribution mechanism is optimized by the use of TMSV states whose degree of squeezing (and hence entanglement) is determined by the working point of the mechanical receivers. A posteriori, this demonstrates that the study in~\cite{mazzolapatern} entails indeed a configuration of optimal entanglement distribution. We then move to the simplest multipartite setting, consisting of a (generally mixed) three-mode driving field guiding the dynamics of three remote and independent optomechanical cavities. We show that, under conditions of fully symmetric multi-mode driving fields, tripartite inseparability is indeed in order as signaled by the violation of the positivity of partial transposition criterion~\cite{giedke}. Our work embodies the first quantitative investigation on the influence that the parameters characterizing a multi-mode driving field have on the performance of mechanical entanglement distribution. Our results allows for the identification of the optimal resources to use at a given optomechanical working point as well as, conversely, the way an optomechanical network should be adjusted in order to fully exploit the entanglement brought about by the driving field. 

The remainder of this manuscript is organized as follows. In Section~\ref{model} we introduce the model used for entanglement distribution among mechanical receivers that are part of simple optomechanical devices and sketch the analytical procedure to determine the stationary reduced mechanical state resulting from the interaction with the driving field. Section~\ref{twomode} discusses the two-mode case and the corresponding efficiency of the entanglement distribution process. This analysis is then extended to a three-site case in Section~\ref{threemode}, where symmetric and genuinely tripartite entangled fields are used in order to get a three-mode inseparable mechanical state. Finally, in Section~\ref{finito} we draw our conclusions and discuss a few open point that we aim at addressing in future. 

As a final remark of this Introduction, it is important to stress that the working points used for our simulations adhere perfectly to the current experimental state of the art, while the sort of optical states used in order to drive our small optomechanical networks are all well within the reach of linear optics experiments aiming at the engineering of continuous-variable states.

\section{Steady-state mechanical entanglement from optomechanical dynamics}
\label{model}

\begin{figure}[t!]
\center\includegraphics[scale=0.38]{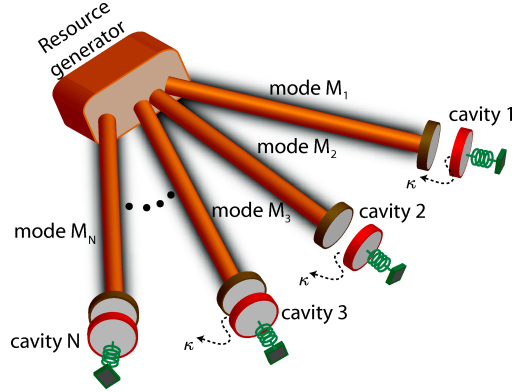}
\caption{Mechanical entanglement distribution scheme. A multi-mode state generator prepares the continuous variable (CV) resource, which is distributed to the nodes of the optomechanical network. The latter consists of remote and non-interacting cavities, each endowed with a light movable mirror of mass effective $m$, mechanical quality factor $Q$ and natural frequency $\omega_m$. Each resonator leaks photons at a rate $\kappa$, has length $L$, free frequency $\omega_c$ and is driven by a quasi-resonant classical pump. The movable mirrors are at equilibrium with their respective environments at temperature $T$.}
\label{scheme}
\end{figure}

We now provide a general description of the mechanical entanglement distribution scheme without limiting the number of sites of the network that we consider. Figure~\ref{scheme} shows a pictorial representation of the experimental configuration at hand, where the optical circuitry needed in order to generate an  optical $N$-mode entangled resource is {\it condensed} in an unspecified ``Resource generator". On the other hand, the mechanical network consists of $N$ remote and mutually independent optomechanical resonators. In order to provide a simple explicit example, we consider the case of Fabry-Perot cavities each endowed with a light mechanical end-mirror, which is one of the configurations currently utilized experimentally~\cite{optoreview} and has been exploited to provide the first experimental evidence of radiation pressure-based passive cooling~\cite{gigan} and demonstrate strong optomechanical coupling~\cite{strong}. Throughout the manuscript, for the sake of simplicity, we assume identical optomechanical nodes. Therefore, each cavity has length $L$, natural frequency $\omega_c$, photon decay rate $\kappa$ and is terminated by a mechanically movable mirror of effective mass $m$, free frequency (under harmonic approximations) $\omega_m$ and mechanical quality factor $Q$. Cavity $j=1,2,..,N$ is driven by both a large-intensity classical field (pump power $\eta$ and frequency $\omega_0$) and mode $M_j$ of the (weak) optical entangled resource. The drivings populate the cavity field, which in turn is coupled, via radiation pressure at bare rate $\chi=\omega_c/L$~\cite{law}, to the mechanical mirror. The free energy of the whole optomechanical network thus reads
\begin{equation}
\label{ham}
\hat{H}=\sum^N_{j=1}\hat H_{j}=\sum^N_{j=1}\left[\hbar(\delta-\chi\hat{q}_j)\hat{a}^\dag_{j}\hat a_j+\frac{1}{2m}\hat p^2_j+\frac12m\omega_m^2\hat q^2_j+\hbar{E}\hat y_j\right],
\end{equation}
where $\delta=\omega_c-\omega_0$ is the cavity-pump detuning, ${E}=\sqrt{\frac{2\kappa\mu}{\hbar\omega_0}}$ is the amplitude of the displacement undergone by the $j^{\rm th}$ cavity field due to the strong pump, $\hat a$ ($\hat a^\dag$) is the corresponding annihilation (creation) operator [with $\hat y_j=\rmi(\hat a^\dag_j-\hat a_j)/\sqrt2$ the out-of-phase quadrature of the cavity field] and $\{\hat q_j,\hat p_j\}$ the position and momentum operator of the $j^{\rm th}$ mechanical oscillator satisfying the canonical commutation relations $[\hat q_j,\hat p_k]=\rmi\hbar\delta_{jk}~(j,k=1,..,N)$. The dynamics induced by this Hamiltonian is captured by the quantum Langevin equations of the whole optomechanical system (we introduce the vector of operators for subsystem $j$ $\hat {O}_j=(\hat q_j,\hat p_j,\hat x_j,\hat y_j)$ with $\hat x_j=(\hat a^\dag_j+\hat a_j)/\sqrt2$)
\begin{equation}
\partial_t\hat{O}_j=\frac\rmi\hbar[\hat H,\hat {O}_j]+{\rm noise~terms}~~~~(j=1,..,N),
\end{equation}
which incorporate the mechanical damping at rate $\gamma_m=\omega_m/Q$, the cavity photon decay and two {\it noise} contributions that deserve a more detailed introduction. The first is the incoherent Brownian dynamics induced by the interaction between the $j^{\rm th}$ mechanical mode and the ensemble of phononic modes excited in the substrate onto which the mechanical oscillator is fabricated~\cite{paternostroNJP}. In principle, this mechanism gives rise to non-Markovian evolutions, as encompassed by the two-time correlation function of the (zero-mean) Langevin force operator $\hat{b}_j$ describing Brownian motion~\cite{giovannetti}
\begin{equation}
\langle\hat b_j(t)\hat b_j(t')\rangle=\frac{\hbar\gamma_m}{2\pi}m\int\omega e^{-i\omega(t-t')}\left[\coth\left(\frac{\hbar\omega}{2k_{\rm B}T}+1\right)\right]d\omega.
\end{equation}
Here $T$ is the temperature of the phononic background modes (with Ohmic spectral density) with which each mechanical oscillator is assumed to be in thermal equilibrium and $k_{\rm B}$ is the Boltzmann constant. Evidently, the noise affecting the mechanical mode is, in general, not delta-correlated. However, for large mechanical quality factors, $\langle\hat b_j(t)\hat b_j(t')\rangle$ becomes approximately delta-correlated, thus reinstating a Markovian description of the associated noise. The validity of such approximation is guaranteed across this paper. 

The second source of noise is due to the open nature of the cavities, which determines the effect of the weak entangled driving fields onto the dynamics of the mechanical oscillators. In fact, in our model, the pumped optomechanical network is affected by an $N$-mode quantum-correlated environment that we use in order to distribute mechanical entanglement. The details of the correlations shared by the environmental modes depend on the form of the state that we decide to use as an entangled resource for the distribution mechanism and will be specified later on in this paper by assigning the appropriate vector of two-time correlation functions ${\bm C}_{\rm in}(t,t')$ (with $t$ and $t'$ two arbitrary instants of time in the system's evolution).

Finally, under the assumption of intense pump fields, we can focus on the dynamics of the quantum fluctuations $\delta\hat O_j$ around the mean value of each operator $\hat O_j$, which are thus expanded as
\begin{equation}
\hat O_j\simeq\overline{O}_j+\delta\hat O_j~~~~~~(j=1,..,N).
\end{equation}
Each $\overline O_j$ that is determined by the steady-state Langevin equations~\cite{paternostroNJP}. It should be stressed that the optomechanical subsystems are mutually independent, so that each light-mechanical mode interaction occurs regardless of similar events taking place at other sites of the network. In this sense, by neglecting the sources of noise, the time-evolution operator of the whole system would read $\hat{U}(t)=\otimes^N_{j=1}e^{-\rmi\hat H_j t}$ with $\hat H_j$ that rules the coupling between mode $M_j$ and oscillator $j$, as specified in~\eref{ham}. The quantum correlations shared by the input noise link the various optomechanical subsystems to each other. 

While details on specific and interesting instances of such correlated environments will be given in the next Sections, here we briefly sketch the way one can track down the evolution of the system. First, the Langevin equations for the quantum fluctuations introduced above can be compactly cast into the vector form $\partial_t[\delta\hat{O}_j(t)]={K}_j\delta\hat{O}_j(t)+\hat{\cal N}_j(t)$ with the dynamical matrix 
\begin{equation}
\label{eqs}
K_j=
\left(\begin{array}{cccc}
0&1/m&0&0\\
-m\omega^2_m&-\gamma_m&\sqrt2\hbar\chi \Re({c_{\rm ss}})&\sqrt2\hbar\chi \Im({c_{\rm ss})}\\
-\sqrt{2}\chi \Im(c_{\rm ss})&0&-\kappa&\Delta\\
\sqrt{2}\chi \Re(c_{\rm ss})&0&-\Delta&-\kappa
\end{array}\right)
\end{equation}
and the vector of noise fluctuation operators $\hat{\cal N}_j=(0,\hat b_{j}(t),\sqrt{2\kappa}\delta\hat x^{\rm in}_j(t),\sqrt{2\kappa}\delta\hat y^{\rm in}_j(t))$, which depends on the Brownian-motion Langevin force and the quadrature fluctuation operators of the input mode $j=1,..,N$. We have introduced the steady-state amplitude of the field in each cavity $c_{\rm ss}={E}/(\kappa+\rmi\Delta)$ and the detuning $\Delta=\delta-\hbar\chi^2 |c_{\rm ss}|^2/(m\omega^2_m)$. It should be noted that, notwithstanding the non-linear character of the light-mechanical mode interaction, the description in terms of quantum fluctuation operators gives back a fully linear model that guarantees the solvability of the problem encompassed by~\eref{eqs}~\cite{entanglement,paternostroNJP,mazzolapatern}. 

\begin{table}[b]
\caption{Parameters used for the simulations run throughout the manuscript (we assume the same values for each optomechanical subsystem). The set of parameters is taken from the experiment in Ref.~\cite{strong}, where a slightly lower input power was used.}
\centering
\begin{tabular}{c c c}
\hline
\hline
Parameter & Symbol & Value \\
\hline
\hline
Mechanical mass & $m$ & 145 ng\\
Mechanical frequency & $\omega_{\rm m}/2\pi$ & 947 KHz\\
Cavity length & $L$ & 25 mm  \\
Input power & ${\eta}$ &  20 mW \\
Cavity-field wavelength & $\lambda$ & 1064 nm \\
Optical damping rate & $\kappa/2\pi$ & 215 KHz \\
Mechanical quality factor & $Q$ & 7000 \\
\hline
\hline
\end{tabular}
\label{tavola}
\end{table}

Here we will be interested in the steady-state regime achieved at long interaction times after all transients induced by the light-mechanical mode interaction have faded away. We thus drop the time derivative in the Langevin equations 
and solve the associated time-dependent set of linear algebraic equations in the Fourier domain. In the following, we will concentrate on the properties of the mechanical system only. We thus report the solution of the Langevin equations for the position and momentum operators of the $j^{\rm th}$ mechanical oscillator, which read
\begin{eqnarray}
\label{soluz1}
\fl
\delta\hat{\cal P}_j&=&\frac{m \omega [\sqrt{2\kappa} \hbar \chi  ({c}_{\rm ss}
  {\delta\hat{\cal C}^{\rm in\dag}_j} (-\Delta +\rmi \kappa +\omega )+{\delta\hat{\cal C}}^{\rm in}_j c^*_{\rm ss}  (\Delta +\rmi \kappa +\omega ))+\rmi\delta{\hat{\cal B}_j}
  (\Delta ^2+(\kappa -\rmi \omega)^2)]}{2{|c_{\rm ss}|^2} \hbar
   \Delta  \chi ^2+m(\rmi{\gamma_{\rm m}} \omega +\omega
   ^2-\omega_{\rm m}^2) \left(\Delta ^2+(\kappa -i \omega
   )^2\right)},\\
   \label{soluz2}
 \fl
\delta\hat{\cal Q}_j&=& \frac{\rmi \sqrt{2\kappa} \hbar \chi  [{c}_{\rm ss} {\delta\hat{\cal C}^{\rm in\dag}_j}
   (-\Delta +\rmi \kappa +\omega )+{\delta\hat{\cal C}^{\rm in}_j} {c}^*_{\rm ss}  (\Delta +\rmi
   \kappa +\omega)]-\delta{\hat{\cal B}_j}[\Delta ^2+(\kappa -\rmi \omega
   )^2] }{2 |{c}_{\rm ss}|^2 \hbar \Delta  \chi ^2  +m(\rmi
  {\gamma_{\rm m}} \omega +\omega ^2-\omega_{\rm m}^2)
  (\Delta ^2+(\kappa -\rmi \omega )^2)}.
   \end{eqnarray}
In the equations above, $\delta\hat{\cal O}_j$ stands for the Fourier-transformed fluctuation operator $\delta\hat O_j$ (the dependence on frequency is understood) and $\delta\hat{x}^{\rm in}_j=(\delta\hat c^{\rm in\dag}_j+\delta\hat c^{\rm in}_j)/\sqrt2$, $\delta\hat{y}^{\rm in}_j=\rmi(\delta\hat c^{\rm in\dag}_j-\delta\hat c^{\rm in}_j)/\sqrt2$. Quite clearly, the dynamics of the $j^{\rm th}$ mechanical oscillator depends crucially on the input noise fluctuation operators: from \eref{soluz1} and~\eref{soluz2} it is evident that any correlation function of the mechanical variables will be determined by the input noise correlators and, in turn, the entanglement that the {\it engineered} environment brings about. In the next Sections we specify the set of input correlators and determine the entanglement that is distributed among the elements of the small mechanical network that we are working on. Finally, throughout our simulations we will use the set of parameters for the optomechanical system given in Table~\ref{tavola}, which match the current experimental state of the art, thus making our proposal very close to feasibility. 

\section{Two-site optomechanical network}
\label{twomode}

Here we restrict our study to the case of $N=2$ sites of the network, which is thus driven by a two-mode quantum correlated state of light. The linearity of the dynamics of the quantum fluctuations ensures that the corresponding map preserves the Gaussian nature of an input optomechanical state. This observation allows us to limit our attention to just this class of continuous-variable states. In fact, Gaussian states are those most easily generated in quantum optics laboratories. Moreover, 
Gaussian resources appear to maximize the efficiency of schemes for the transfer of entanglement from continuous to discrete-variable systems, when compared to {\it experimentally accessible} non-Gaussian states such as photon-subtracted or added fields~\cite{paternostrovari}. Finally, as a key assumption of our model is that each mechanical oscillator is initially in thermal equilibrium with its own background of phononic modes, we take the elements of the mechanical network as initialized in thermal states with mean occupation number $\nbar_j=[{\rme}^{\frac{\hbar\omega_m}{k_{\rm B}T}}-1]^{-1}$. The Gaussian-preserving nature of the problem allows us to use the formalism of covariance matrices: given an $N$-mode systems described by the $2N$-long set of quadrature operators $\hat{Q}=\{\hat{q}_1,\hat{p}_1,..,\hat{q}_N,\hat{p}_N\}$, we define the covariance matrix elements $({\bm\sigma})_{kl}=\langle\hat Q_k\hat Q_l+\hat Q_l\hat Q_k\rangle/2~(k,l=1,2,..,2N)$. The covariance matrix of a physical state should satisfy the Heisenberg-Robertson uncertainty principle ${\bm\sigma}+\rmi\Omega_N/2\ge0$ with $\Omega_N=\oplus^N_{j=1}\rmi\sigma_y$ the so-called symplectic matrix  and $\sigma_y$ the $y$-Pauli matrix. The Gaussian state of the system is determined by the assignment of ${\bm\sigma}$. 

\begin{figure}
\center{{\bf (a)}\hskip5cm{\bf (b)}}\hskip5cm{\bf (c)}
\center{\includegraphics[scale=0.34]{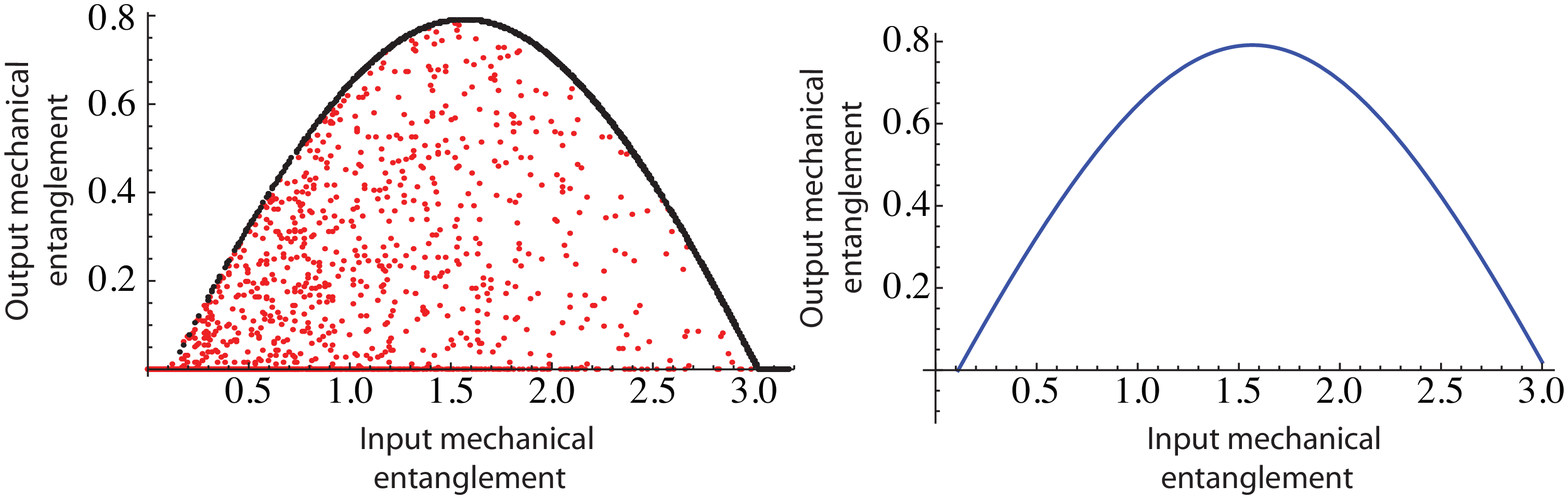}~~~~\includegraphics[scale=0.38]{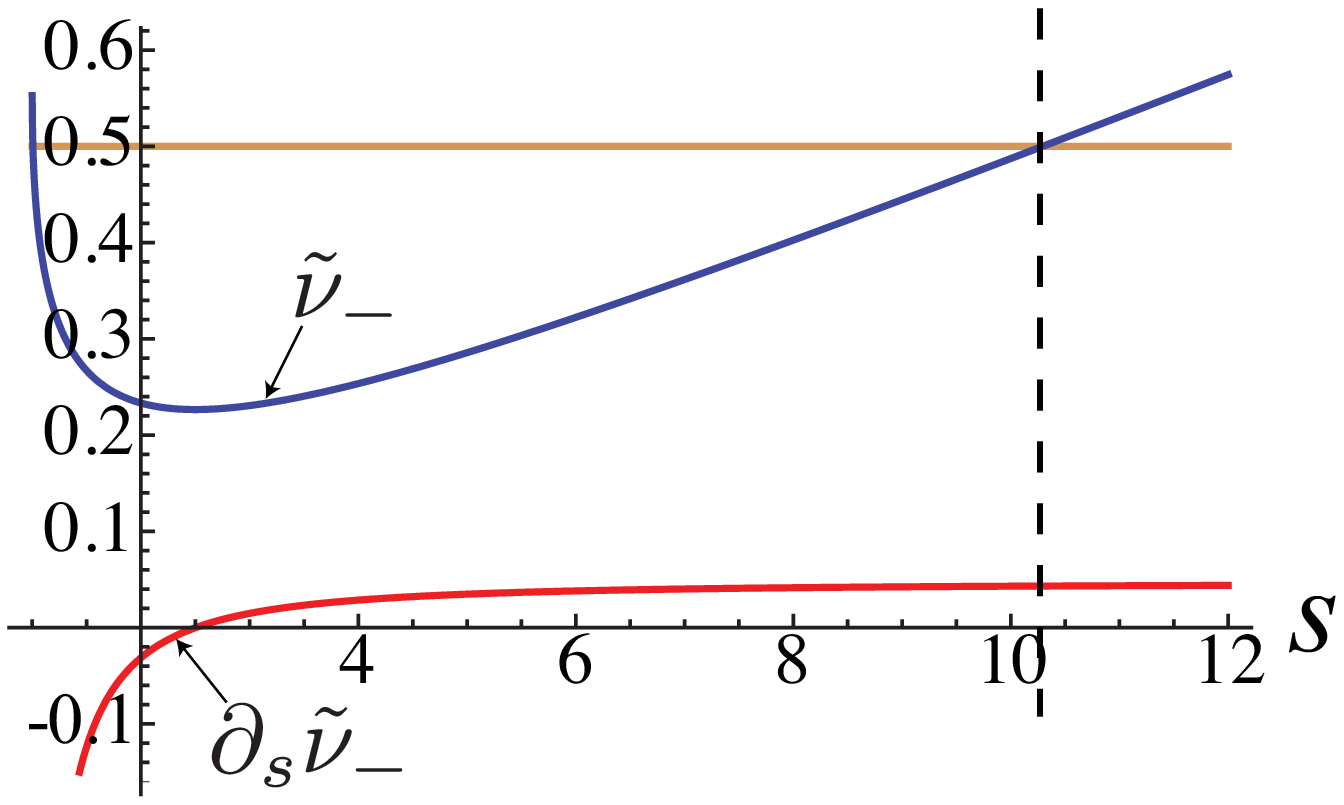}}
\caption{{\bf (a)} Output mechanical entanglement against input optical one for a symmetric two-mode resource with identical single-mode variances ($d=0$) and variable degree of entanglement. The inner red dots represent the distributed mechanical entanglement corresponding to $10^4$ randomly taken mixed input states. On the other hand, the black points on the boundary stand for the mechanical entanglement achieved when $5000$ pure ($g=1$) symmetric input states (picked up randomly) are used. {\bf (b)} The numerical bound that is highlighted in panel {\bf (a)} enjoys an analytical form (whose derivation is sketched in the body of the paper) which is displayed for comparison with the numerics.  {\bf (c)} We study $\tilde\nu_-$ and its derivative with respect to $s$ for a pure, symmetric two-mode resource state. We find $\tilde\nu_-{<}1/2$ for values of $s$ up to the the dashed vertical lines. At $s=2.501$, $\tilde\nu_-$ has a minimum (the entanglement has a maximum).}
\label{conbound}
\end{figure}

We now need a general description of the two-mode resource that we are going to employ. The covariance matrix of any (pure or mixed) two-mode Gaussian state can be put in the so-called {\it standard form} parameterized as~\cite{ai}
\begin{equation}
\label{cm}
{\bm\sigma}=\frac{1}{2}
\left(\begin{array}{cccc}
a&0&c_+&0\\
0&a&0&c_-\\
c_+&0&b&0\\
0&c_-&0&b
\end{array}\right)
\end{equation}
with 
\begin{eqnarray}
a=s+d,\quad b=s-d, \quad c_\pm=\frac{\sqrt{(f_d-h_d)^2-4g^2}\pm\sqrt{(f_s-h_d)^2-4g^2}}{4\sqrt{s^2-d^2}}
\end{eqnarray}
 and $f_{x}=4x^2+(g^2+1)(\lambda-1)/2~(x=d,s)$. The state corresponding to ${\bm\sigma}$ is physical and entangled for $s\ge{1},|d|\le s-1, 2|d|+1\le{g}\le{2s-1}$ and $\lambda\in[-1,1]$. Here, $s$ and $d$ set the local entropy of each mode, $g$ determines the global purity of the state, while the degree of entanglement shared by the two modes grows with $\lambda$. The assignment of the input optical covariance matrix ${\bm\sigma}_{\rm opt}$ specifies the set of two-times correlators ${\bm C}_{\rm in}(t,t')$ needed to solve the mechanical dynamics. Indeed, a covariance matrix such as \eref{cm} is obtained by taking an optical state having the only non-zero correlators  as 
 \begin{eqnarray}
 \fl
{\bm C}_{\rm in}(t,t')&=&(\langle\delta\hat c^{\rm in\dag}_1(t)\delta\hat c^{\rm in}_1(t')\rangle, \langle\delta\hat c^{\rm in}_1(t)\delta\hat c^{\rm in\dag}_1(t')\rangle,\langle\delta\hat c^{\rm in\dag}_2(t)\delta\hat c^{\rm in}_2(t')\rangle,\\
 \fl
&& \langle\delta\hat c^{\rm in}_2(t)\delta\hat c^{\rm in\dag}_2(t')\rangle,\langle\delta\hat c^{\rm in}_1(t)\delta\hat c^{\rm in}_2(t')\rangle, \langle\delta\hat c^{\rm in\dag}_1(t)\delta\hat c^{\rm in\dag}_2(t')\rangle)\\
 \fl
 &=&\left(a-\frac{1}2,a+\frac12,b-\frac{1}2,b+\frac12,\rme^{-\rmi\omega_m(t-t')}\frac{(c_+{-}c_-)}2,\rme^{\rmi\omega_m(t-t')}\frac{(c_+{+}c_-)}2\right)\delta(t-t').
 \end{eqnarray}
 Physically, this corresponds to taking a weak driving field whose modes have frequency $\omega_0+\omega_m$~\cite{mazzolapatern}. 
 The covariance matrix ${\bm\sigma}_{\rm mech}$ resulting from the assignment of a ${\bm\sigma}_{\rm opt}$ equal to \eref{cm} maintains the standard form and reads
 \begin{equation}
 \label{out}
 {\bm\sigma}_{\rm mech}=\left(
\begin{array}{cccc}
 \langle\delta\hat Q_1^2\rangle&0& \langle\delta\hat Q_1\delta\hat Q_2\rangle&0\\
 0& \langle\delta\hat P_1^2\rangle&0& \langle\delta\hat P_1\delta\hat P_2 \rangle\\
 \langle\delta\hat Q_1\delta\hat Q_2\rangle&0&\langle\delta\hat Q_2^2\rangle&0\\
  0&\langle\delta\hat P_1\delta\hat P_2 \rangle&0&\langle\delta\hat P_2^2\rangle
  \end{array}\right),
 \end{equation}
 where each element is calculated using~\eref{soluz1} and~\eref{soluz2} and then transforming back to the time domain~\cite{mazzolapatern}. We have generated a random sample of input two-mode states by taking parameters uniformly distributed across their ranges of variation, which are in turn determined by setting an upper bound to the possible values of $s$. Correspondingly, we have calculated the output entanglement distributed between the mechanical oscillators. This has been quantified using logarithmic negativity~\cite{wernervidal}, which for Gaussian states is defined as ${\cal L}_{\cal N}=\max[0,-\ln2\tilde\nu_-]$, where $\tilde\nu_-=\min{\rm eig}|\rmi\Omega_2\tilde{\bm\sigma}|$ is the minimum symplectic eigenvalue of the covariance matrix $\tilde{\bm\sigma}=P_{1|2}{\bm\sigma}P_{1|2}$ associated with the partially transposed state of the mechanical network~\cite{ai} ($P_{1|2}={\rm diag}[1,1,1,-1]$ is the matrix that, by inverting the sign of momentum of the mechanical oscillator $2$, realizes partial transposition at the level of covariance matrices~\cite{simon}). The results are displayed in Figure~\ref{conbound} {\bf (a)}, where we compare the input optical entanglement (determined calculating the logarithmic negativity of the two-mode input state) and the corresponding output mechanical entanglement associated with (\ref{out}) and calculated at the detuning $\Delta=\omega_m$ that maximizes it. Although the range of entanglement of the input states extends uniformly from 0 to more than 3, mechanical entanglement can only be distributed when the input one is larger than a non-zero threshold determined by the working parameters at the optomechanical level and not the input field's properties. We have also found that the system favors symmetric input states: the number of entangled mechanical states obtained when considering input states characterized by $d=0$ is much larger than the analogous number for asymmetric states. This results from the uniformity of the parameters within the optomechanical network and is an effect that can be cured by matching the optical asymmetry with an equal mechanical one. Most importantly, the entangled distributed in the mechanical network is constrained to stay below a curve that maximizes the degree of output entanglement at assigned values of the input one. We have found that the optical resource giving rise to such boundary should have $d=0$ and $g=\lambda=1$, i.e. it should be a pure, symmetric state (thus equivalent to a TMSV) with covariance matrix
 \begin{equation}
 {\bm\sigma}_{\rm opt}=\frac12
\left(\begin{array}{cccc}
s&0&\sqrt{s^2-1}&0\\
0&s&0&-\sqrt{s^2-1}\\
\sqrt{s^2-1}&0&s&0\\
0&-\sqrt{s^2-1}&0&s
\end{array}\right).
\end{equation}
The behavior of the boundary suggested by the numerical exploration is confirmed by an exact analytical approach which we sketch here. First, we write the input optical entanglement $\epsilon$ in terms of the parameter $s$ characterizing the optimal resource as
\begin{equation}
\epsilon=-\frac12\ln|{{1-2{s}^2+2s \sqrt{{s}^2-1}}}|\rightarrow s=\cosh\epsilon.
\end{equation} 
This expression for $s$ can then be replaced in the analytical one for $\tilde\nu_-$, the minimum symplectic eigenvalue of the partially transposed covariance matrix of the mechanical network. While it is possible to get an exact closed expression for $\tilde\nu_-$ as a function of all the parameters of the optomechanical setting, this is too lengthy to be reported here. In Figure~\ref{conbound} {\bf (b)} we show such analytic bound for the parameters used in panel {\bf (a)}. We defer an analysis of the behavior of such expression against some of the key parameters of our system to later.

Therefore, a TMSV embodies {\it the} ideal resource for the distribution of entanglement in the mechanical network, thus proving that the analysis in~\cite{mazzolapatern} is optimal. Moreover, we can determine the value of $s$ that, at the optimal detuning $\Delta=\omega_{\rm m}$, maximizes the transferred entanglement, thus identifying the optimal input resource for the distribution scheme. To do this, we look for the solution of the equation $\partial_s\tilde\nu_-(s)=0$ at $g=\lambda=1$ and for a symmetric resource. Figure~\ref{conbound} {\bf (c)} shows the trend followed by both $\tilde\nu_-$ and its derivative with respect to $s$. At $s=2.501$, the symplectic eigenvalue of the partially transposed mechanical state achieves its minimum: again, the analytic expression for such solution is too uninformative to be shown here. 

\begin{figure}[t]
\center{{\bf (a)}\hskip5.5cm{\bf (b)}}
\center{\includegraphics[scale=0.44]{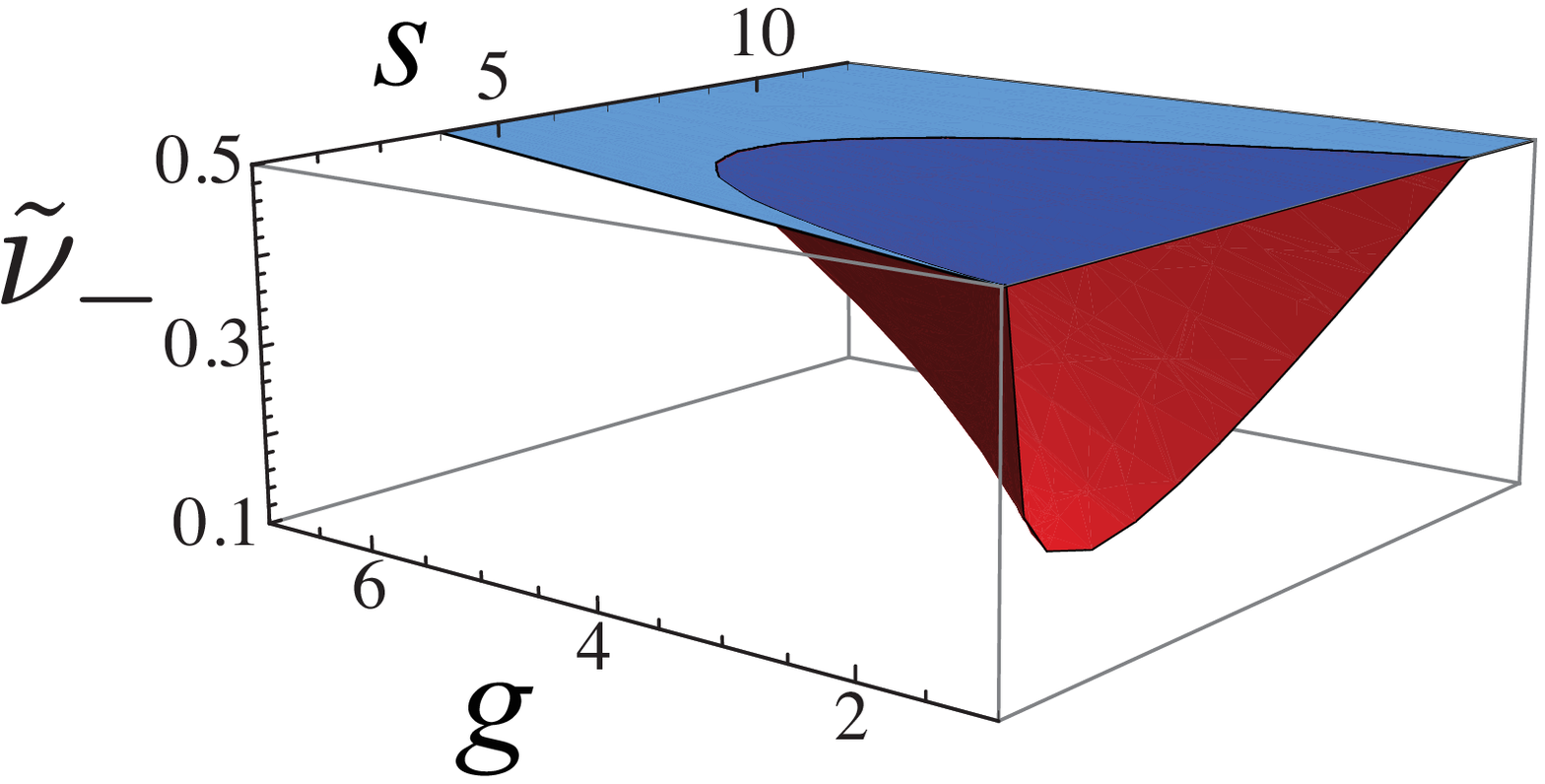}~~\includegraphics[scale=0.45]{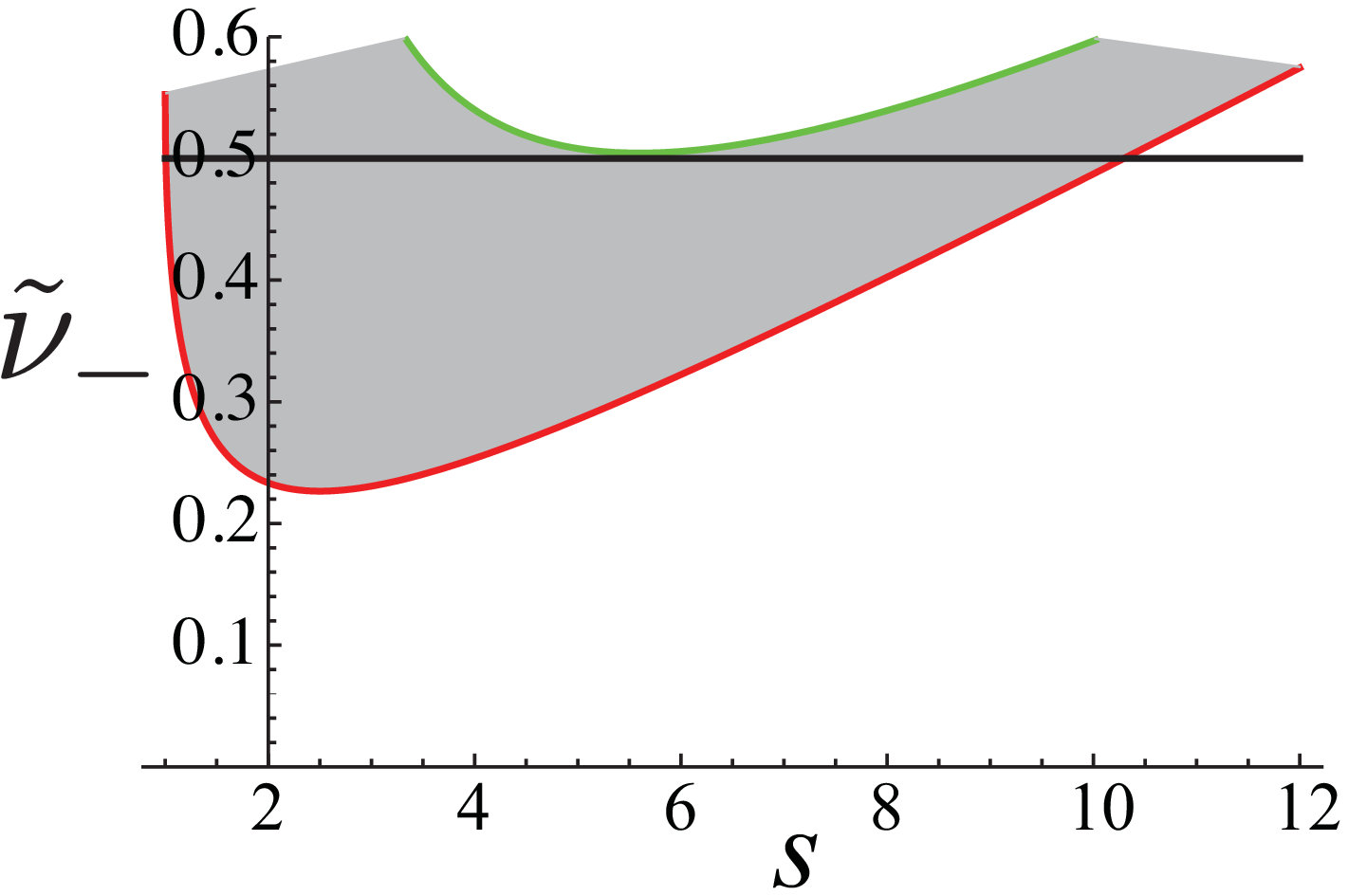}}
\caption{{\bf (a)} Minimum symplectic eigenvalue of the partially transposed covariance matrix of the mechanical system studied against the single-mode variance $s$ and the global state purity $g$ of the input resource, for $d=0$ and $\lambda=1$. Whenever $\tilde\nu_->1/2$, mechanical entanglement disappears. {\bf (b)} The function $\tilde\nu_-(g,s)$ is asymmetric: the larger $g$, the larger the value of $s$ at which entanglement is maximized. The range of $s$ within which entanglement is distributed to the mechanical network is maximum for $g=1$ (lower red boundary) and shrinks as $g$ grows. At $g=5.01$ (corresponding to the upper green boundary) all mechanical entanglement disappears. }
\label{simplettico}
\end{figure}

While, in line with expectations, we have checked that an increasing mass and/or mechanical damping rate reduce the amount of entanglement between the mechanical systems, the effects of the deviations from ideality of the resource are yet to be assessed. This is relevant given that it is likely that the optical resource produced in a quantum laboratory would be mixed (due, for instance, to residual thermal nature at the two-mode state generation stage). We thus now address the effects that a non-unit purity of the two-mode optical resource has on the performance of our scheme. In order to do this we have considered the range of values of $(s,g)$ for a symmetric optical input state (with $\lambda=1$) that leaves $\tilde\nu_-<1/2$. As shown in Figure~\ref{simplettico} {\bf (a)}, the scheme tolerates a considerable degree of mixedness of the input resource: only at $g=5.01$ (corresponding to a state purity of 0.2) and for the optomechanical parameters stated in Table~\ref{tavola}, the distributed entanglement disappears. The value of $s$ that optimizes the entanglement distribution increases with $g$, making the function $\tilde\nu_-(g,s)$ asymmetric with respect to $s$. This is seen in Figure~\ref{simplettico} {\bf (b)}, where we highlight the region of the $(g,s)$ space within which mechanical entanglement is found.

\begin{figure}[t]
\center{{\bf (a)}\hskip5cm{\bf (b)}}
\center{\includegraphics[scale=0.45]{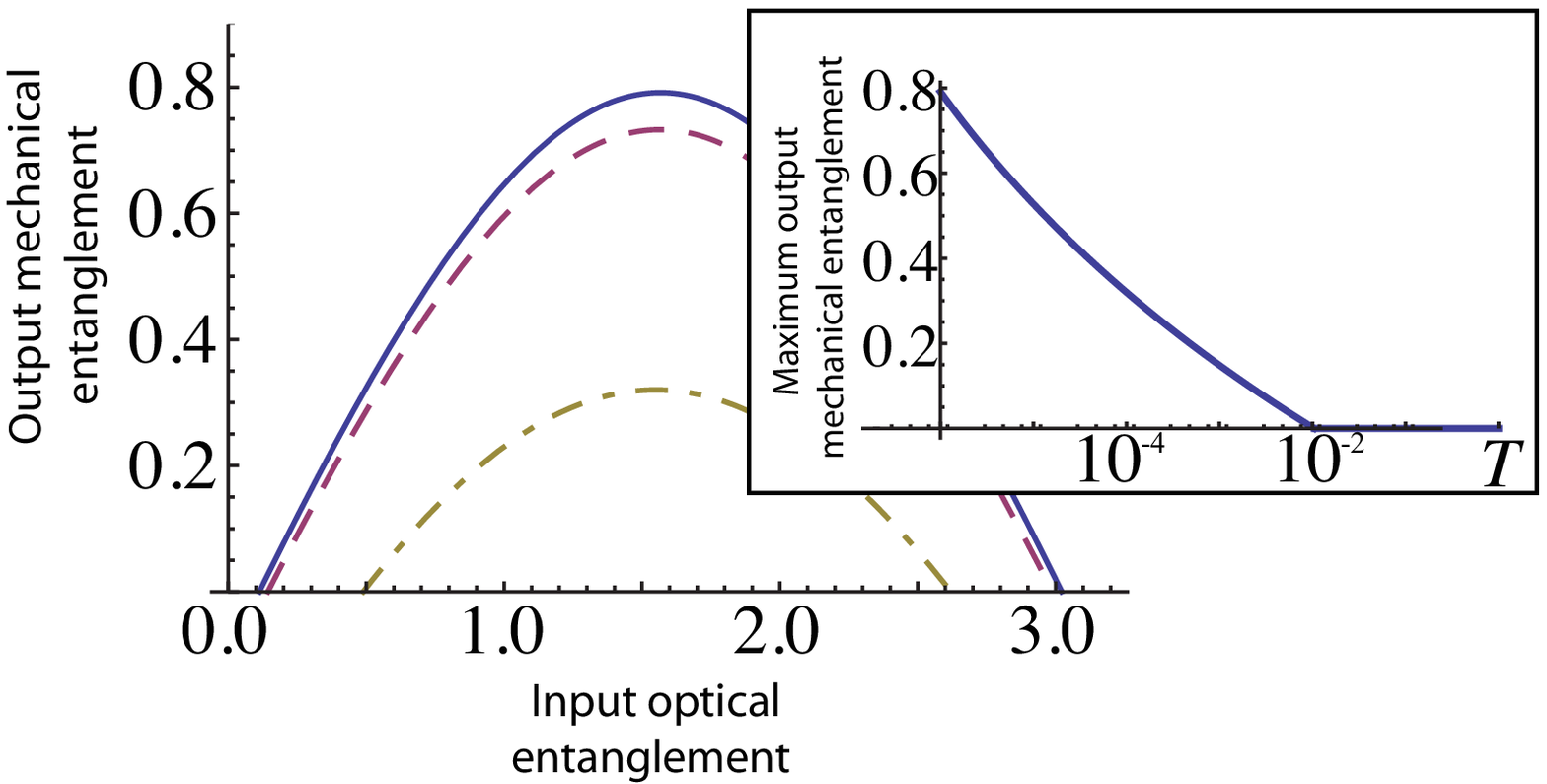}~~~~~~\includegraphics[scale=0.45]{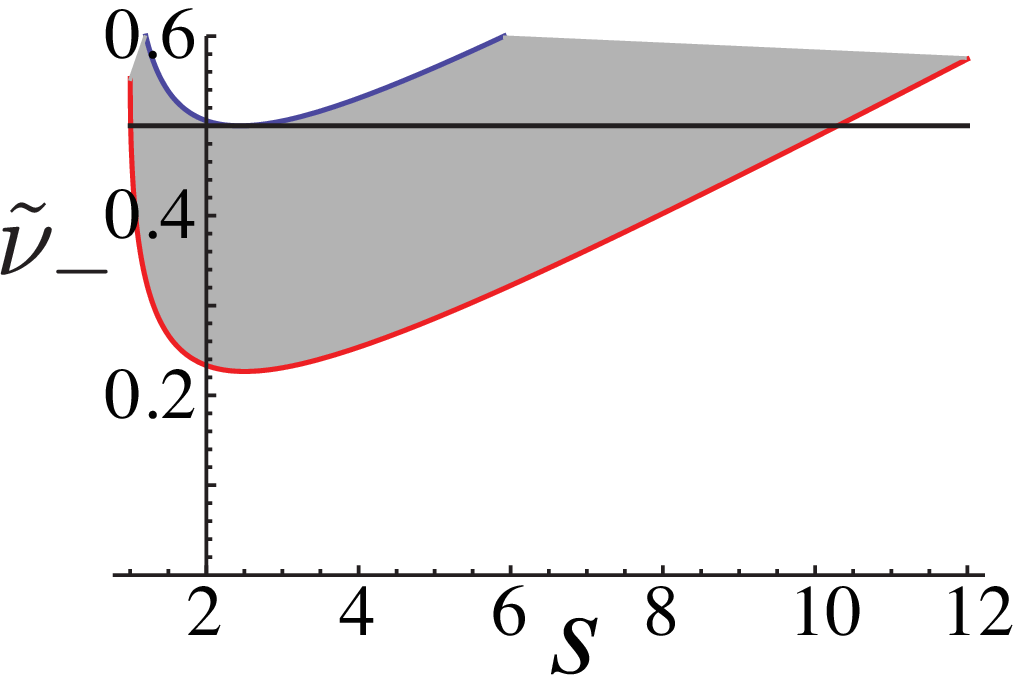}}
\caption{{\bf (a)} Output mechanical entanglement against input optical one for input driving fields characterized by $d=0$, $g=1$ and $\lambda=1$. From top to bottom, the curves correspond to $T=10^{-6}$K (top-most solid line), $T=10^{-4}$K (dashed curve), $T=10^{-3}$K (dot-dashed curve). Inset: we show the behaviour of the maximum output mechanical entanglement against the temperature of the mechanical modes (in logarithmic scale). {\bf (b)} We plot $\tilde\nu_-$ against $s$ for $g=1$. The lower (solid) curve corresponds to $T=10^{-6}$K, while the upper one shows the case corresponding to $T=2.01\times10^{-2}$K. Any situation having $T$ between such extremal values gives rise to $\tilde\nu_-$'s falling within gray-shadowed region. }
\label{tempe}
\end{figure}

Another important parameter to consider is the initial temperature $T$ of the mechanical systems. How {\it warm} should the network be in order to jeopardize the entanglement distribution mechanism? The process appears to be robust against thermal effects. Figure~\ref{tempe} {\bf (a)} shows that the decrease of the maximum value of the output mechanical entanglement is quasi exponential and very flat at small temperatures (the drop in the maximum entanglement from $T=10^{-6}$K to $10^{-4}$K is only quite marginal. Notice that the horizontal axis is in logarithmic scale). Entanglement persists up to $T=0.02$K, showing that even macroscopically occupied thermal states (at $T=0.01$K the mean thermal occupation number of a mode at frequency $\omega_{\rm m}$ is larger than $10^3$) can accommodate the inputted entanglement [cf. Figure~\ref{tempe} {\bf (a)} (inset) and {\bf (b)}].



\section{Three-site optomechanical network}
\label{threemode}

We now assume that a three-mode optical driving field is prepared to drive the dynamics of a three-site optomechanical network. While the principles of the dynamics at the single optomechanical site are the same as those for the two-site case, the description of the optical resource needs some care. The covariance matrix of a pure symmetric three-mode state is parameterized as~\cite{ai}
\begin{equation}
{\bm\sigma}=\frac12\left(
\begin{array}{cccccc}
a_1&0&c^+_{12}&0&c^+_{13}&0\\
0&a_1&0&c^-_{12}&0&c^-_{13}\\
c^+_{12}&0&a_2&0&c^+_{23}&0\\
0&c^-_{12}&0&a_2&0&c^-_{23}\\
c^+_{13}&0&c^+_{23}&0&a_3&0\\
0&c^-_{13}&0&c^-_{23}&0&a_3
\end{array}\right)
\end{equation}
with 
\begin{eqnarray}
c^\pm_{ij}&=&\frac{\sqrt{[({a_i}-{a_j})^2-({a_k}-1)^2]
  [({a_i}-{a_j})^2-({a_k}+1)^2]}}{4 \sqrt{{a_i}{a_j}}}\\
   &\pm&\frac{\sqrt{[({a_i}+{a_j})^2-({a_k}-1)^2]
  [({a_i}+{a_j})^2-({a_k}+1)^2]}}{4 \sqrt{{a_i}{a_j}}}.
\end{eqnarray}
In order to ensure physicality of the corresponding state, the parameters $a_j$'s should satisfy the triangular inequality
\begin{equation}
|{a_i}-{a_j}|+1\leq{a_k}\leq
  {a_i}+{a_j}-1~~~(i\neq j\neq k=1,2,3~{\rm with}~a_j\ge1).
\end{equation}
The analysis presented in Section~\ref{twomode} has revealed that pure and symmetric optical states enable better performances of our scheme, at least for the two-site case. We now conjecture that such condition extends to the three-site situation and restrict our attention to the case of $a_j=a~(\forall j)$, therefore considering the 
covariance matrix of the pure three-mode optical state
\begin{equation}
\label{cov3}
{\bm\sigma}_{\rm opt}=\frac12
\left(
\begin{array}{cccccc}
 a & 0 &{x^+} & 0 &{x^+} & 0 \\
 0 & a & 0 &{x^-} & 0 &{x^-} \\
{x^+} & 0 & a & 0 &{x^+} & 0 \\
 0 &{x^-} & 0 & a & 0 &{x^-} \\
{x^+} & 0 &{x^+} & 0 & a & 0 \\
 0 &{x^-} & 0 &{x^-} & 0 & a
\end{array}
\right)
\end{equation}
with $x^\pm=\left( a^2-1 \pm \sqrt{1 - 10 a^2 + 9 a^4}\right)/(8 a)$. Such a conjecture has been verified {\it a posteriori} by taking covariance matrices with lower degrees of symmetry (such as cases where the variances of only two out of three modes are identical) and checking that the corresponding performances of the protocol were inferior to those of the fully symmetric case. The state with covariance matrix given by Eq.~(\ref{cov3}) is genuinely tripartite entangled according to the measure given by the so-called minimum residual Gaussian contangle~\cite{ai2} ${\cal E}_{\rm res}$. The Gaussian contangle ${\cal E}_{a|b}$ between the pair of subsystems $a$ and $b$ is defined as the convex roof (performed all the over pure Gaussian states with covariance matrix ${\bm\sigma}_{\rm Gauss}$ strictly smaller that the covariance matrix of the state at hand) of the squared logarithmic negativity~\cite{ai} and is a proper entanglement monotone. For a set of modes $\mu=\{i,j,k,l,..,z\}$, the Gaussian contangle satisfies the monogamy constraint ${\cal E}_{i|jkl..z}-\sum_{\mu=\{j,k,l,..,z\}}{\cal E}_{i|\mu}\ge0$, together with the analogous relations obtained by permuting the mode indices at will. With this, for three modes $(i,j,k)$, ${\cal E}_{\rm res}$ is defined as
\begin{equation}
{\cal E}_{\rm res}=\min_{\hat\Pi}\hat\Pi[{\cal E}_{i|jk}-{\cal E}_{i|j}-{\cal E}_{i|k}]
\end{equation}
with $\hat\Pi$ the operator that permutes the indices $\{i,j,k\}$. The residual entanglement of the symmetric state with covariance matrix $2{\bm\sigma}_{\rm opt}$ (the factor 2 is used to uniform the notation to the formalism of~\cite{ai2}) can be computed analytically as ${\cal E}_{\rm res}={\rm arcsinh}^2[\sqrt{a^2-1}]-(1/2)\ln^2[2a^2+4ax^-]$. Moreover, the state has non null entanglement between any two modes, growing (but not indefinitely) with $a$. Driving fields of this type are within the grasp of current experimental possibilities~\cite{ai}. 

The corresponding mechanical covariance matrix is the extension of \eref{out} to three mechanical modes. Throughout this Section we maintain the very same values of the optomechanical parameters given in Table~\ref{tavola}. Needless to say, the output covariance matrix ${\bm\sigma}_{\rm mech}$ is physical as far as ${\bm\sigma}_{\rm mech}+\rmi\Omega_3/2\ge0$ (which is equivalent to the condition on the symplectic spectrum $\min{\rm eig}|\rmi\Omega_3{\bm\sigma}_{\rm mech}|>1/2$). As for entanglement, given the tripartite nature of the mechanical network at hand, the situation is more involved than in the two-site case. Indeed, we can distinguish the case of bipartite 1-vs-1-mode entanglement, achieved by tracing out any one of the three mechanical modes, from the 1-vs-2-mode entanglement, which we will use to infer the tripartite inseparable nature of the mechanical state.  

\begin{figure}[b]
\center{{\bf (a)}\hskip5.5cm{\bf (b)}\hskip5.5cm{\bf (c)}}
\includegraphics[scale=0.35]{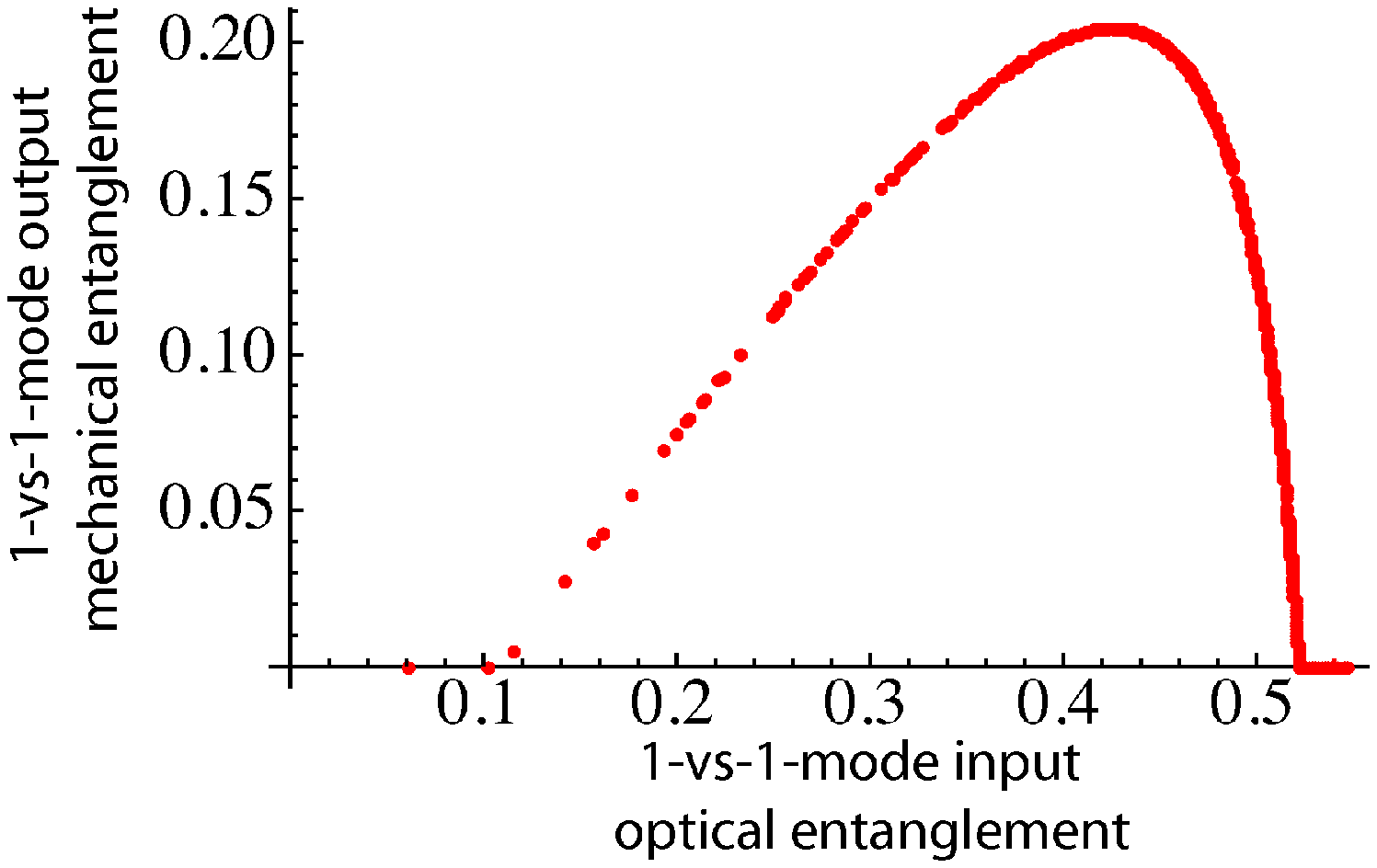}~~\includegraphics[scale=0.35]{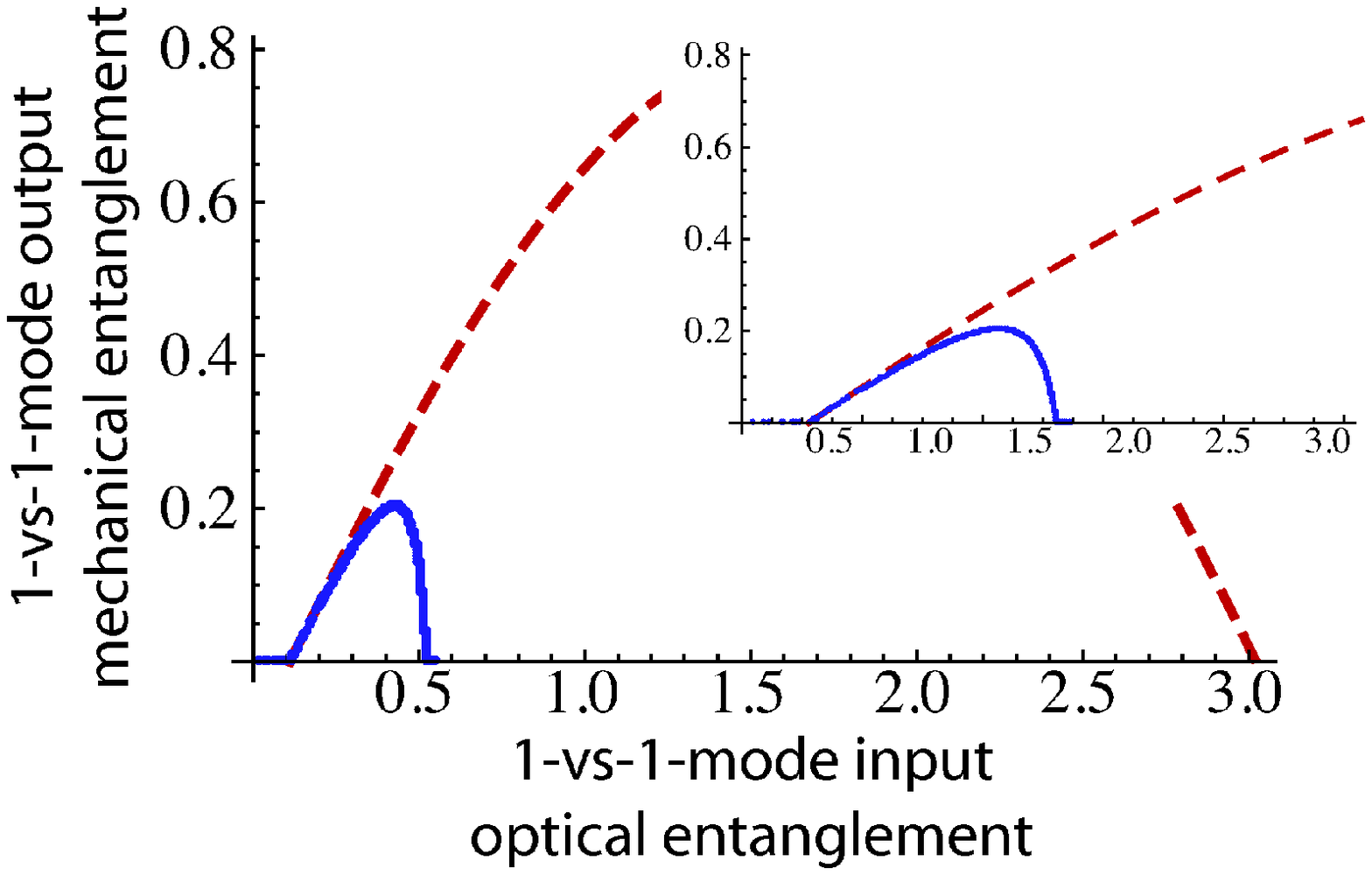}~~\includegraphics[scale=0.35]{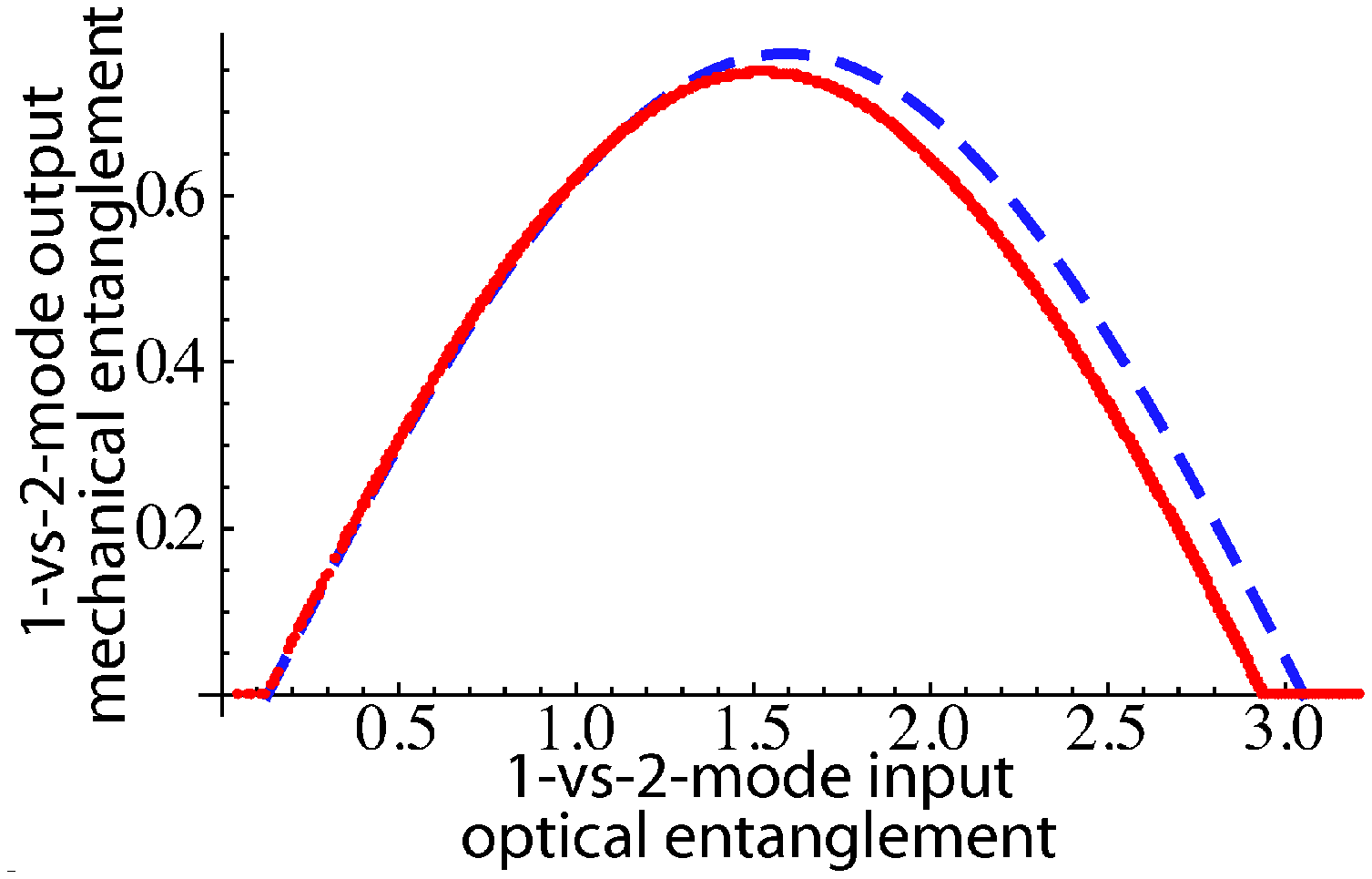}
\caption{{\bf (a)} One-vs-one-mode entanglement for a three-site optomechanical network pumped by a pure and symmetric three-mode optical state. We plot the output mechanical entanglement against the input optical one for $3000$ randomly picked values of $a$. {\bf (b)} Comparison between one-vs-one-mode entanglement (i.e. the one involving any two modes out of the three-site mechanical network) and the two-mode mechanical entanglement as determined in Section~\ref{twomode}. The first is contained fully within the second. {\bf (c)} One-vs-two-mode entanglement (solid line) for a three-site network for the very same states used in panel {\bf (a)}. The curve is fully contained within the area below the optimally distributed two-mode mechanical entanglement (dashed line).}
\label{tremodi}
\end{figure}

Let us start with the first instance, which implies the trace over the degrees of freedom of one of the modes of our mechanical network. Needless to say, the enforced symmetry of the optical input state guarantees full symmetry of the mechanical one, thus making all the two-mode reductions identical to each other. Formally, this is equivalent to consider only the {\it blocks} in the mechanical covariance matrix ${\bm\sigma}_{\rm mech}$ containing the degrees of freedom of the modes that are not traced out. Entanglement is then calculated by using the very same formal apparatus described in Section~\ref{twomode} and we can characterize the output mechanical entanglement of the two-mode reduced states versus the analogous quantity for the input optical one [cf. Figure~\ref{tremodi} {\bf (a)}]. In analogy with the case of two-site networks, we observe both the non-monotonic behavior of the output mechanical entanglement and the existence of a threshold in the input optical entanglement for the process to be successful. However, at variance with what has been observed in Section~\ref{twomode}, the curve of the output entanglement is asymmetric with respect to the input one. 
Needless to say, as the reduced two-site system is mixed, we expect the distribution performances in this case to be {\it contained} within the optimal boundary embodied by a TMSV. This is shown in Figure~\ref{tremodi} {\bf (b)} and its inset. 

Symmetry is fully reinstated when one looks at the 1-vs-2-mode splitting of the network. Entanglement in such configuration (which is again invariant under permutations of the mode indices due to the overall symmetry of the three-mode mechanical covariance matrix) is detected by the violation of the Heisenberg-Robertson condition by the partially transposed covariance matrix $P_{i|jk}{\bm\sigma}_{\rm mech}P_{i|jk}$, where $P_{i|jk}$ is the matrix that inverts the sign of momentum of mode $i$, thus implementing the partial transposition with respect to such element of the mechanical network. Entanglement in such one-vs-two-mode split is then quantified as ${\cal L}_{\cal N}=\max[0,-\ln2\tilde\nu_-]$ with $\tilde\nu_-=\min{\rm eig}|\rmi\Omega_3P_{i|jk}{\bm\sigma}P_{i|jk}|$ (as the partially transposed covariance matrix has only one symplectic eigenvalue that can be smaller than $1/2$) and is plotted, against the input optical entanglement, in Figure~\ref{tremodi} {\bf (c)}. Not surprisingly, the similarity with Figure~\ref{conbound} {\bf (b)} is only apparent and the one-vs-two-mode entanglement is strictly contained within the one-vs-one-mode entanglement.

With this at hand, we can proceed to the qulitative characterization of the tripartite entanglement within the system at hand. Following the classification by Giedke {\it et al.}~\cite{giedke}, a tripartite Gaussian state whose one-vs-two-mode bipartitions are inseparable is tripartite inseparable. Looking at the results shown in Figure~\ref{tremodi} {\bf (c)}, this occurs, in our system and for the set of optomechanical parameters  at hand, for quite a large set of input optical states, thus showing that multipartite entanglement used as a resource in our protocol is indeed able to generate tripartite entangled states of the mechanical network, in a sense signaling the success of the protocol itself. 

Yet, a point that remains to be addressed is the quantification of the genuine tripartite entanglement witin the full mechanical state. As the state of the mechanical network is overall mixed, the calculation should be performed with care. Without entering unnecessary details related to the rather technical approach to the calculation of the residual genuine tripartite entanglement in the state of a system of bosonic modes
we mention that the results of such calculations show that genuine tripartite entanglement is indeed transferred to the mechanical network, albeit in an inefficient manner (the average tripartite mechanical entanglement being two order of magnitude smaller than the corresponding input one, in general). Yet, this is sufficient for the scopes of this investigation, which aimed at providing a formal analysis of the distribution performance in the simplest instance of multipartite optomechanical network.

\section{Conclusions}
\label{finito}

In this paper we have addressed extensively the problem of entanglement distribution in an optomechanical network that is driven by a multipartite entangled optical resource. By using the most general parameterization of (generally mixed) Gaussian optical states, we have found the conditions under which two-site mechanical entanglement is maximally distributed identifying the role that symmetries and local/global state purity of the optical resource play in this process. We have demonstrated the efficiency of the protocol, which can be implemented using state of the art parameters and the possibility to distribute genuine tripartite mechanical entanglement.

Many questions remain to be addressed, for instance related to the ability of the protocol to pass general quantum correlations that go beyond quantum entanglement. This is an interesting opportunity in light of the preliminary analysis performed in~\cite{mazzolapatern}, where such quantum correlations have been shown to be more robust than entanglement to the increase of temperature in the mechanical setup. Moreover, it would be interesting to find whether optical postselection are able to improve the distribution mechanism, giving rise to more substantial genuinely multipartite mechanical entanglement. Finally, we intend to perform a classification of the distributed mechanical entanglement at assigned values of the global and marginal purity, along the lines of some of the studies in~\cite{paternostrovari}, so as to check if a hierarchy of mechanical states arises from the distribution process. 

\section{Acknowledgements}

MP is supported by the UK EPSRC through a Career Acceleration Fellowship and the ``New Directions for EPSRC Research Leaders" initiative (EP/G004759/1). LM acknowledges the EU under a Marie Curie IEF Fellowship.

\end{document}